# Effect of van der Waals interactions on the stability of SiC polytypes

Sakiko Kawanishi[†*] and Teruyasu Mizoguchi

Institute of Industrial Science, The University of Tokyo, 4-6-1 Komaba, Meguro-ku, Tokyo 153-8505, Japan

E-mail: s-kawa@tagen.tohoku.ac.jp

**Abstract**

Density functional theory calculations with a correction of the long-range dispersion force, namely the van der Waals (vdW) force, are performed for SiC polytypes. The lattice parameters are in good agreement with those obtained from experiments. Furthermore, the stability of the polytypes in the experiments, which show 3C-SiC as the most stable, are reproduced by the present calculations. The effect of the vdW force on the electronic structure and the stability of polytypes are discussed. We observe that the vdW interaction is more sensitive to the cubic site than the hexagonal site. Thus, the influence of the vdW force increases with decreasing the hexagonality of the polytype, which results in the confirmation that the most stable polytype is 3C-SiC.

---

[†] Present address: Institute of Multidisciplinary Research for Advanced Materials, Tohoku University, 2-1-1 Katahira, Aoba-ku, Sendai, Miyagi 980-8577, Japan.

Silicon carbide (SiC) has attracted great interest as a promising widegap material for use in power devices with low on-resistance because of its high breakdown field. SiC is also well known as a material with polytypism, such as 3C-, 2H-, 4H-, 6H-, 15R-SiC and so forth. Among the SiC polytypes, 4H-SiC has been desired for application in power devices because of its wide band gap and relatively isotropic electron mobility. Although the supply of 4H-SiC wafers with 6 inch size has begun, a large problem remains regarding the difficulty in selective growth of 4H-SiC. So far, many investigations have been performed for understanding the stability of the SiC polytypes experimentally[1,2] and theoretically.[3-7] However, the experimentally observed stability of the SiC polytypes,[1] which show 3C-SiC is more stable at less than 1873 K compared with 4H-, 6H- and 15R-SiC, has never been reproduced by a first principles density functional theory (DFT) calculation. In the DFT calculation, 4H-, 6H- or 15R-SiC is estimated to be the most stable among the typical polytypes.[4-6] The discrepancy has been proposed to arise from the entropic term and crystal growth conditions owing to the estimated total energy difference between 3C- and 4H-SiC, which denotes that the internal energy difference at 0 K is within a few meV.[4] However, the long-range dispersion force, that is, the van der Waals (vdW) force, has never been included in the previous DFT calculations for SiC polytypes despite the observation that vdW often affects the electronic structure of covalent materials.[8]

Recently, including the vdW force in DFT calculations, namely the DFT-D calculation, has been performed to describe the accurate electronic structure of two-dimensional layered structures and molecular crystals.[8-12] Although SiC is a highly covalent material, the effect of the vdW interaction on the stability of SiC polytypes has not been discussed so far. Herein, the stability of SiC polytypes are evaluated by the DFT-D calculation. The lattice parameters of the typical SiC polytypes are determined, and the effect of the vdW force on electronic structure and phase stability are discussed.

Among the SiC polytypes, 2H-, 4H-, 6H-, 15R- and 3C-SiC were considered in this study, because of their importance in engineering applications and their availability. Their crystal structures are shown in Fig. 1. The first principles calculations were performed using the plane wave based projector augmented wave (PAW) method based on DFT, implemented in Vienna Ab-initio Simulation Package (VASP). The exchange correlation interaction was considered by generalized gradient approximation (GGA) as well as local density approximation (LDA). To take the long-range dispersion force into account, Tkatchenko–Scheffler (TS)[10] correction to GGA, which is developed based on the DFT-D2 method,[11] was applied to the calculations. The dispersion energy ($E_{disp}$) within the TS method is formally identical with that of the DFT-D2 method and is expressed using empirical parameters as follows,

$$E_{disp} = -\frac{1}{2}\sum_{i=1}^{N_{at}}\sum_{j=1}^{N_{at}}\sum_{L}{'}\frac{C_{6ij}}{r_{ij,L}^6}f_{d,6}(r_{ij,L}), \quad (1)$$

where the summations are over all atoms ($N_{at}$) and all translations of the unit cell ($L = (l_1, l_2, l_3)$), the prime indicates that $i \neq j$ for $L = 0$, $C_{6ij}$ denotes the dispersion coefficient for the atom pair $ij$, $r_{ij,L}$ is the distance between atom $i$ located in the reference cell $L = 0$ and atom $j$ in the cell $L$, and the term $f(r_{ij})$ is a damping function whose role is to scale the force field such as to minimize contributions from interactions within typical bonding distances. Here, the dispersion coefficient and damping function in the TS method depend on the ground state electron-density.[10] The plane wave cut-off energy was set to 500 eV, and the mesh size of k-points in an irreducible Brillouin zone was set to 15 × 15 × 15 for 2H- and 3C-SiC, 15 × 15 × 5 for 4H- and 6H-SiC, and 15 × 15 × 2 for 15R-SiC. Here, the convergence of the calculation condition was carefully confirmed. For instance, the difference of the total energy of 4H-SiC calculated by 15 × 15 × 5 and 20 × 20 × 8 k-mesh was confirmed to be better than 0.1 meV / Si-C pair. Also, the total energy difference between 4H- and 3C-SiC calculated by 800 eV cut-off energy was confirmed to be better than 0.1 meV / Si-C pair compared with that calculated by 500 meV.

*Insert Fig. 1 here.*

The calculated lattice parameters for the *a* and *c* axes of 2H-, 4H-, 6H-, 15R- and 3C-SiC are shown in Table I. In addition, Jagodzinski's *hk* notation[13] and the hexagonality of each polytype are also provided. Here, the *hk* notation represents a description of the stacking structure of close-packed planes to distinguish hexagonally (*h*) and cubically (*k*) stacked layers. Hexagonality is the ratio of the hexagonal sites to the total number of sites, that is, $h / (h + k)$. Results from the present study are compared with other calculations[3] and experiments.[14-20] The lattice constants obtained using the DFT-GGA and DFT-LDA without consideration of the vdW force slightly differ from the experimental values. Hence, DFT-GGA and DFT-LDA calculations show larger and smaller values than the experiments, respectively, which is usually observed in the GGA and LDA calculations. However, the lattice constants are well reproduced for all the polytypes by including the vdW force into the calculation. The experimentally obtained lattice constants for the *c* axis per Si-C site indicate the increasing tendency with increasing hexagonality, which are also reproduced in the present calculation.

*Insert Table I here.*

Figure 2 shows the total energy difference of each polytype relative to 3C-SiC. The most stable polytype without consideration of the vdW force was calculated to be 4H-SiC, and the total energy of 3C-SiC was a few meV higher than that of 4H-SiC, which is the same tendency as the other theoretical calculations.[3-7] However, when the vdW force was taken into account in the GGA, the tendency of the stability largely changed. The total energies per Si-C pair for 2H-, 4H-, 6H- and 15R-SiC are 15, 2.7, 1.3 and 1.9 meV higher than that for 3C-SiC by considering the vdW force. 3C-SiC was estimated to be the most stable polytype, which agrees with the experimental polytype stability at less than 1873 K.[1]

*Insert Fig. 2 here.*

To assess the effect of the vdW force on the stability of SiC polytypes, the distribution of the charge density along the $c$ axis, which represents the character of the hexagonal structure, was investigated. The charge density from the C atom to the neighboring Si atom for hexagonal ($h$) and cubic ($k$) sites of 4H-SiC with or without the vdW force based on GGA are shown in Fig. 3(a) and (b), and their difference is shown in Fig. 3(c). As can be seen from Fig. 3(a) and (b), C has a larger charge density than Si, indicating that electron transfer from Si to C occurred. By subtracting the charge density without the vdW force from that with the vdW force, it is found that the amount of electron transfer is larger when the vdW force is considered. Furthermore, it is also found that the amount of the charge transfer becomes much larger in the $k$-site than that in the $h$-site by considering the vdW interaction (Fig. 3(c)). From Eq. (1), the vdW force becomes larger when the distance between the atoms is short, and the distance for the $k$-site is estimated to be 0.4% shorter than that for the $h$-site. This indicates that the electron density around the $k$-site is more sensitively changed by the vdW force than that around the $h$-site. The same trend was also observed for the other polytypes.

Figure 4 shows the estimated vdW energy together with the experimentally obtained lattice constants for the $c$ axis per Si-C site[14-20] for each polytype. The vdW force on the $k$-site has a greater effect compared with that on the $h$-site, which arises from the shorter distance between atoms along the $c$ axis in the $k$-site. This can be also seen experimentally in the lattice constant, as shown in Fig. 4. Hence, the vdW energy was found to highly affect the lower hexagonality polytype. An almost linear relation between the vdW energy and hexagonality was estimated, and a larger effect of the vdW force of about 10 meV was estimated for 3C-SiC, compared with that for 2H-SiC. As shown in Fig. 2, the difference of the total energy of each polytype was of the same order, resulting in the change of the polytype stability of SiC by the effect of the vdW force. It was thus clarified that the vdW force is of great importance to the stability of the SiC polytypes.

*Insert Figs. 3 and 4 here.*

In summary, the DFT calculations including a correction for the vdW force for SiC polytypes enabled the precise estimation of the lattice parameters. Furthermore, the stabilities of polytypes, where 3C-SiC is the most stable, was reproduced by including the vdW force in the calculation because of the higher effect of the vdW interaction on the *k*-site compared with that on the *h*-site.

Finally, we know that effect of entropy has to be considered to understand the stability of the polytypes. However, our calculation clearly suggests that the vdW interaction plays an important role for the stability of SiC polytypes. Consideration of the vibrational entropy is a next subject for this study, and it will be reported in elsewhere.

**Acknowledgement**

This research was partly supported by Grants-in-Aid (Grant Nos. 26820334, 15J11838, 26249092, 26630302, 25106003) from the Japan Society for the Promotion of Science.

References

1) Y. Inomata, M. Mitomo, Z. Inoue, and H. Tanaka, Yogyo-Kyokai-Shi **77**, 130 (1969) [in Japanese].

2) W.F. Knippenberg, Philips Res. Repts. **18**, 161 (1963).

3) K. Kobayashi and S. Komatsu, J. Phys. Soc. Jpn. **81**, 024714 (2012).

4) C. H. Park, B. H. Cheong, K. H. Lee, and K. J. Chang, Phys. Rev. B **49**, 4485 (1994).

5) P. Käckell, B. Wenzien, and F. Bechstedt, Phys. Rev. B **50**, 10761 (1994).

6) K. Karch, P. Pavone, W. Mindi, O. Schütt, and D. Strauch, Phys. Rev. B **50**, 17054 (1994).

7) S. Limpijumnong and W. R. L. Lambrecht, Phys. Rev. B **57**, 12017 (1998).

8) F. Ricci and G. Profeta, Phys. Rev. B **87**, 184105 (2013).

9) S. Grimme, J. Comput. Chem. **27**, 1787 (2006).

10) A. Tkatchenko and M. Scheffler, Phys. Rev. Lett. **102**, 6 (2009).

11) J. Klimeš, D. R. Bowler, and A. Michaelides, Phys. Rev. B **83**, 195131 (2011).

12) M. Dion, H. Rydberg, E. Schröder, D. C. Langreth, and B. I. Lundqvist, Phys. Rev. Lett. **92**, 246401 (2004).

13) H. Jagodzinski, Acta Cryst. **2**, 201 (1949).

14) H. Schulz and K. H. Thiemann, Solid State Commun. **32**, 783 (1979).

15) R.F. Adamsky and K.M. Merz, Z. Kristallogr. **111**, 350 (1959).

16) R. W. G. Wyckoff, Crystal Structure (Wiley, New York, 1963).

17) M. Stockmeier, R. Müller, S.A. Sakwe, P.J. Wellmann, and A. Magerl, J. Appl. Phys. **105**, 033511 (2009).

18) Y.M. Tairov and V.F. Tsvetkov, Progress in controlling the growth of polytypic crystals,

Pergamon Press, Oxford, New York Vol. 7 (1983), p. 111.

19) A. H. Gomes de Mesquita, Acta Cryst. **23**, 610 (1967).

20) Physics of Group-IV Elements and III–V Compounds, ed. O. Madelung, M. Schulz, and M. Weiss (Springer, Berlin, 1982) Landolt–Bo¨rnstein, New Series, Group III, Vol. 17, Pt. a.

21) A. R.Verma and P.Krishna, Polymorphism and Polytypism in Crystals (Wiley, New York, 1966).

22) A. J. C.Wilson and E.Prince, International Table for Crystallography (Kluwer, Dordrecht, 2004) Vol. C, Chap. 9.2, p. 744.

**Figure captions**

Fig. 1. Crystal structures of 2H-, 3C-, 4H-, 6H- and 15R-SiC. Edge of the unit cell of each structure is described as black solid lines. Gray and yellow circles represent Si and C, respectively. The stacking sequences along the *c*-axis for each polytype are also described by *hk* notation[13] on the left side and ABC notation[21,22] on the right side.

Fig. 2. Total energy of 2H-, 4H-, 6H- and 15R-SiC relative to 3C-SiC.

Fig. 3. Charge density of 4H-SiC from the C atom to the Si atom with or without consideration of the vdW force for (a) the *h*-site and (b) the *k*-site, and (c) the difference of charge density for each site.

Fig. 4 VdW energy and lattice constants along the *c* axis per Si-C pair for each polytype.

Table I. Lattice constants of 2H-, 4H-, 15R-, 6H- and 3C-SiC.

| Polytype | hk notation | Hexagonality (%) | | a (Å) | c (Å) | c / p (Å) |
|---|---|---|---|---|---|---|
| **2H-SiC** | $(h)_2$ | 100 | GGA | 3.091 | 5.073 | 2.537 |
| | | | LDA | 3.058 | 5.018 | 2.509 |
| | | | GGA(vdW) | 3.074 | 5.059 | 2.529 |
| | | | Exp. | 3.079,[14] 3.076[15] | 5.053,[14] 5.048[15] | 2.527,[14] 2.524[15] |
| | | | GGA [3] | 3.050 | 5.006 | 2.503 |
| **4H-SiC** | $(hk)_2$ | 50 | GGA | 3.095 | 10.128 | 2.532 |
| | | | LDA | 3.061 | 10.014 | 2.504 |
| | | | GGA(vdW) | 3.076 | 10.085 | 2.521 |
| | | | Exp. | 3.073,[16] 3.079[17] | 10.052,[16] 10.082[17] | 2.513,[16] 2.520[17] |
| | | | GGA [3] | 3.052 | 9.993 | 2.498 |
| **15R-SiC** | $(hkkhk)_3$ | 40 | GGA | 3.094 | 37.980 | 2.532 |
| | | | LDA | 3.060 | 37.562 | 2.504 |
| | | | GGA(vdW) | 3.076 | 37.800 | 2.520 |
| | | | Exp. [18] | 3.080 | 37.70 | 2.513 |
| **6H-SiC** | $(hkk)_2$ | 33 | GGA | 3.095 | 15.185 | 2.531 |
| | | | LDA | 3.061 | 15.024 | 2.504 |
| | | | GGA(vdW) | 3.077 | 15.113 | 2.519 |
| | | | Exp. | 3.080,[17] 3.081[19] | 15.12,[17] 15.12[19] | 2.520,[17] 2.520[19] |
| | | | GGA [3] | 3.053 | 14.980 | 2.497 |
| **3C-SiC** | $(k)$ | 0 | GGA | 3.097 | 7.587 | 2.529 |
| | | | LDA | 3.063 | 7.503 | 2.501 |
| | | | GGA(vdW) | 3.079 | 7.542 | 2.514 |
| | | | Exp. [20] | 3.083 | 7.552 | 2.517 |
| | | | GGA [3] | 3.055 | 7.482 | 2.494 |

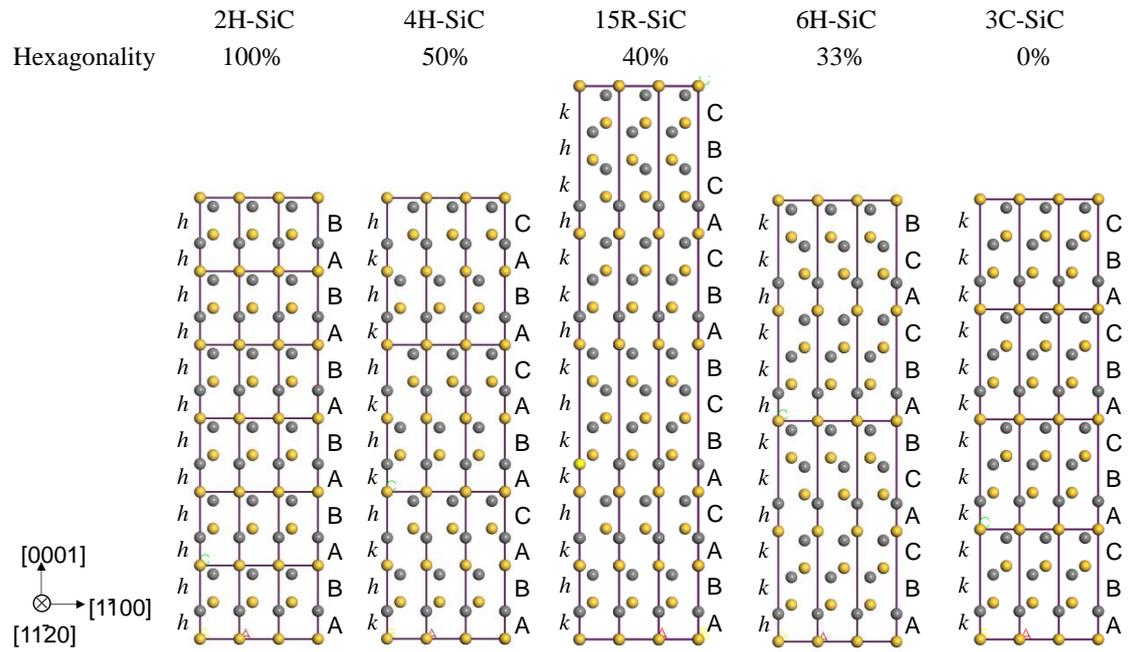

Fig. 1. Crystal structures of 2H-, 3C-, 4H-, 6H- and 15R-SiC. Edge of the unit cell of each structure is described as black solid lines. Gray and yellow circles represent Si and C, respectively. The stacking sequences along the *c*-axis for each polytype are also described by *hk* notation[13] on the left side and ABC notation[21, 22] on the right side.

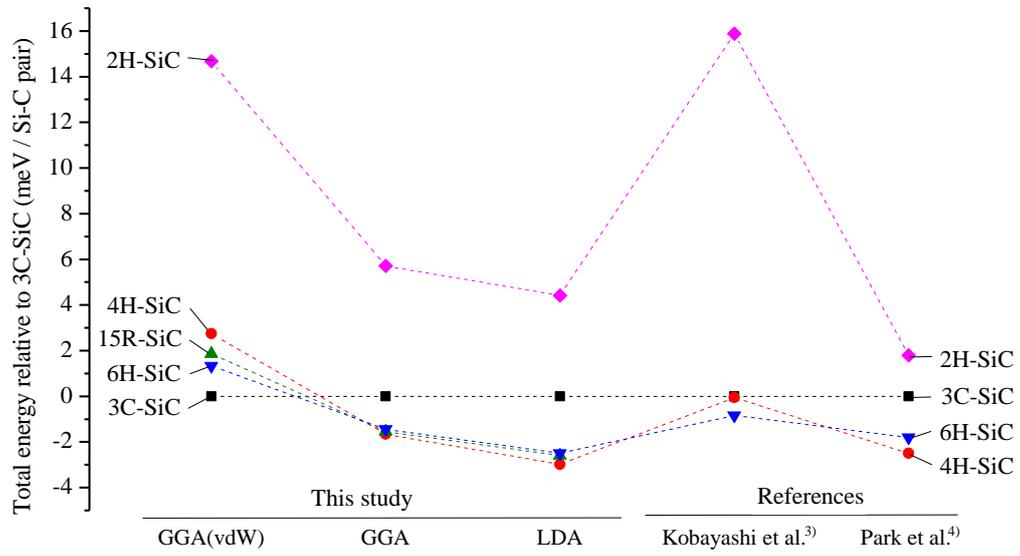

Fig. 2. Total energy of 2H-, 4H-, 6H- and 15R-SiC relative to 3C-SiC.

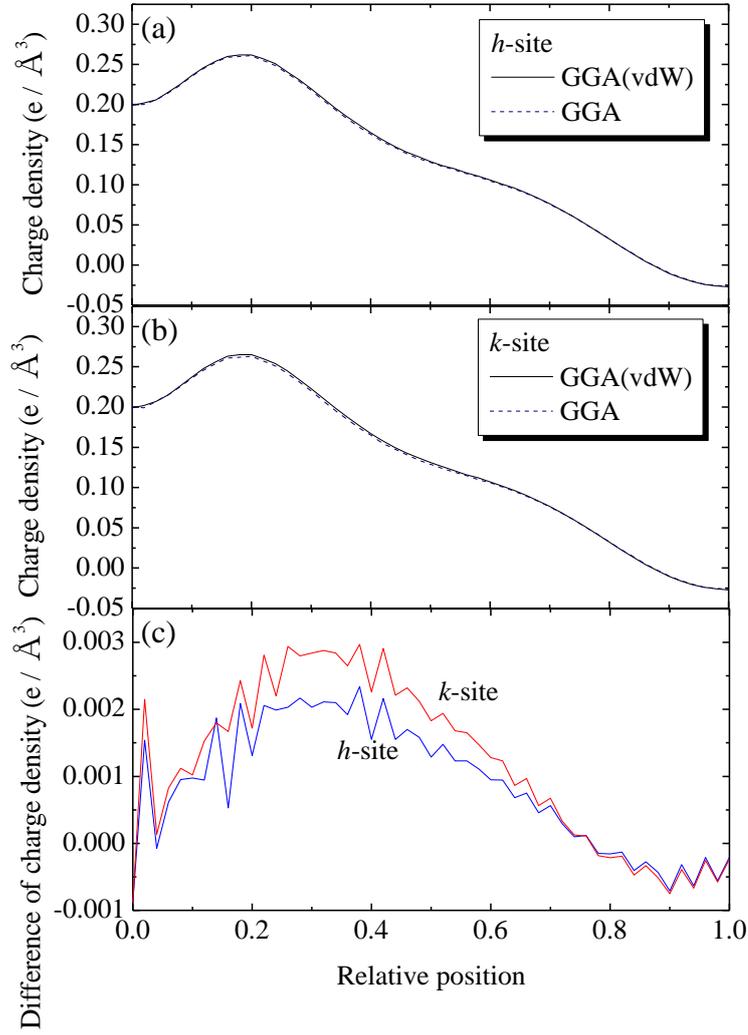

Fig. 3. Charge density of 4H-SiC from the C atom to the Si atom with or without consideration of the vdW force for (a) the *h*-site and (b) the *k*-site, and (c) the difference of charge density for each site.

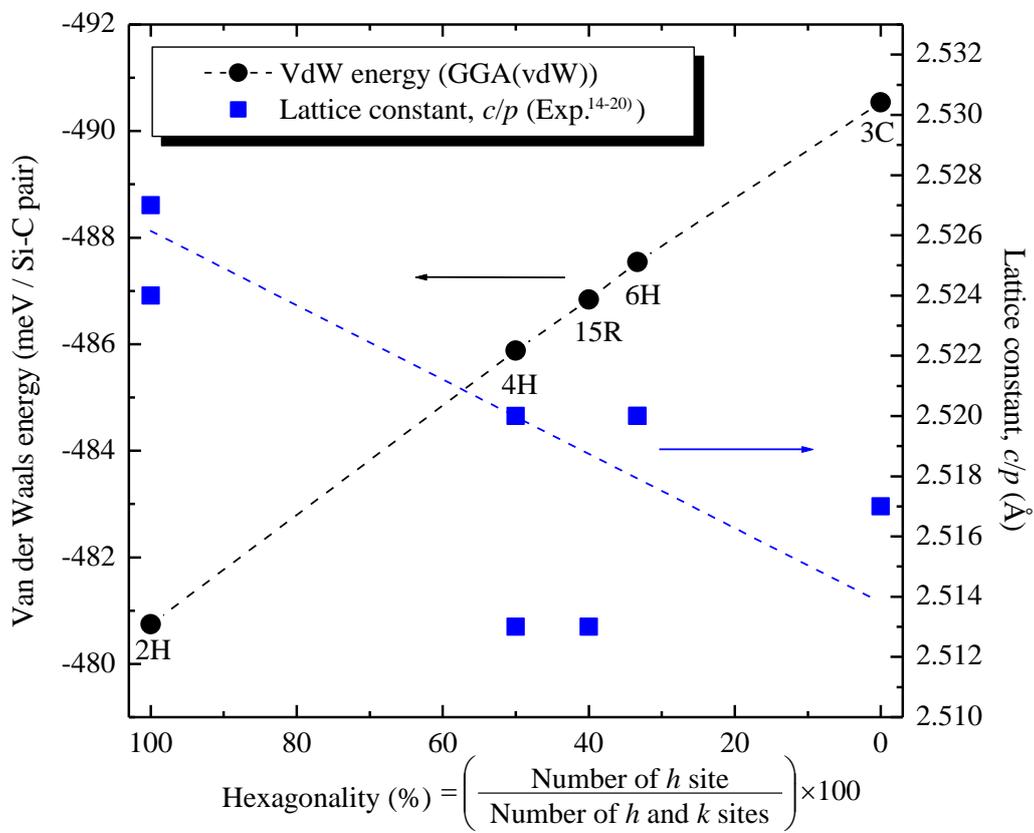

Fig. 4 VdW energy and lattice constants along the *c* axis per Si-C pair for each polytype.